\documentclass[12pt,letter]{article}
\pdfoutput=1
\usepackage{graphicx, epsfig, color,cite}
\usepackage{amsmath}
\usepackage{amssymb}
\usepackage{float}
\usepackage{caption,subcaption,graphicx}
\usepackage{hyperref}

\def\eslt{\not\!\!\!{E_T}}
\def\to{\rightarrow}

\def\bi{\begin{itemize}}
\def\ei{\end{itemize}}

\def\tchi{\tilde\chi}

\def\tst{\tilde t}

\def\tg{\tilde g}

\def\alt{\lesssim}
\def\agt{\gtrsim}
\def\be{\begin{equation}}  
\def\ee{\end{equation}}  
\def\bea{\begin{eqnarray}}  
\def\eea{\end{eqnarray}}

\begin{document}
\begin{titlepage}
\begin{flushright}
OU-HEP-230101
\end{flushright}

\vspace{0.5cm}
\begin{center}
  {\Large \bf Detecting heavy neutral SUSY Higgs bosons\\
    decaying to sparticles at the high-luminosity LHC
    }\\
\vspace{1.2cm} \renewcommand{\thefootnote}{\fnsymbol{footnote}}
{\large Howard Baer$^{1,2}$\footnote[1]{Email: baer@ou.edu },
Vernon Barger$^2$\footnote[2]{Email: barger@pheno.wisc.edu},
Xerxes Tata$^3$\footnote[3]{Email: tata@phys.hawaii.edu} and
Kairui Zhang$^2$\footnote[3]{Email: kzhang89@wisc.edu}
}\\ 
\vspace{1.2cm} \renewcommand{\thefootnote}{\arabic{footnote}}
{\it 
$^1$Homer L. Dodge Department of Physics and Astronomy,
University of Oklahoma, Norman, OK 73019, USA \\[3pt]
}
{\it 
$^2$Department of Physics,
University of Wisconsin, Madison, WI 53706 USA \\[3pt]
}
{\it 
$^3$Department of Physics and Astronomy,
University of Hawaii, Honolulu, HI 53706 USA \\[3pt]
}

\end{center}

\vspace{0.5cm}
\begin{abstract}
\noindent
In supersymmetry (SUSY) models with low electroweak naturalness
(natSUSY), which have been suggested to be the most likely version of SUSY to emerge
from the string landscape, higgsinos are expected at the few hundred GeV
scale whilst electroweak gauginos inhabit the TeV scale. For TeV-scale
heavy neutral SUSY Higgs bosons $H$ and $A$, as currently required by
LHC searches, then the dominant decay modes of $H,\ A$ are into gaugino
plus higgsino provided these decays are kinematically open. The light
higgsinos decay to soft particles so are largely invisible whilst the
gauginos decay to $W$, $Z$ or $h$ plus missing transverse energy
($\eslt$). Thus, we examine the viability of $H,\ A\to W+\eslt$,
$Z+\eslt$ and $h+\eslt$ signatures at the high luminosity LHC (HL-LHC)
in light of large Standard Model (SM) backgrounds from (mainly)
$t\bar{t}$, $VV$ and $Vh$ production (where $V=W,\ Z$). We also examine
whether these signal channels can be enhanced over backgrounds by
requiring the presence of an additional soft lepton from the decays of
the light higgsinos.  We find significant regions in the vicinity of
$m_A\sim 1-2$ TeV of the $m_A$ vs. $\tan\beta$ plane which can be probed
at the high luminosity LHC using these dominant signatures by HL-LHC at
$5\sigma$ and at the 95\% confidence level (CL).

\end{abstract}
\end{titlepage}

\section{Introduction}
\label{sec:intro}

An advantage to searching for (R-parity conserving)
supersymmetry\cite{Baer:2006rs,Drees:2004jm,Baer:2020kwz} (SUSY) via
heavy Higgs boson production at the CERN Large Hadron Collider (LHC) is
that, instead of having to pair produce new states of matter, one may
singly produce some of the new $R$-even states directly via $s$-channel
resonances. In the Minimal Supersymmetric Standard Model (MSSM), this
means direct production of the heavy scalar and pseudoscalar Higgs
bosons, $H$ and $A$, respectively. Indeed, LHC measurements of the
properties of the light Higgs boson $h$ so far have shown it to be
nearly Standard Model (SM)-like\cite{Cadamuro:2019tcf}.  This situation
is expected in the decoupling regime where the heavy SUSY Higgs bosons,
and possibly also many sparticles, are well beyond current LHC
reach.\footnote{ A very SM-like light Higgs boson can also be obtained
  in the alignment
  regime\cite{Gunion:2002zf,Craig:2013hca,Carena:2013ooa} where the new
  Higgs bosons, $H$ and $A$, need not be so heavy.}  

The most stringent LHC Run 2 limits on heavy Higgs bosons have been
obtained by the ATLAS\cite{ATLAS:2020zms} and CMS\cite{CMS:2022goy}
collaborations by searching for $H,\ A\to\tau\bar{\tau}$ with $\sim 139$
fb$^{-1}$ of integrated luminosity. These heavy Higgs search limits are
presented in the $m_A$ vs. $\tan\beta$ plane within the so-called
$m_h^{125}$ scenario as proposed by Bagnaschi {\it et al.} in
Ref. \cite{Bagnaschi:2018ofa}. 
In the $m_h^{125}$ benchmark scenario,
most SUSY particles are taken to be at or around the 2 TeV scale, with a
SUSY $\mu$ parameter at $\mu =1$ TeV. This ensures that SUSY particles
only slightly affect the heavy Higgs searches, and that the dominant $H$
and $A$ decay modes are into SM particles. The ATLAS exclusion contour,
which is also shown in Fig.~\ref{fig:disc_excl} below, shows that the
Higgs decoupling limit with a heavy SUSY spectrum is now a likely
possibility, particularly since LHC Run 2 limits with $\sim 139$
fb$^{-1}$ of integrated luminosity seem to require gluino masses
$m_{\tg}\agt 2.2$ TeV\cite{ATLAS:2020syg,CMS:2019zmd} and top squark
masses $m_{\tst_1}\agt
1.2$~TeV\cite{ATLAS:2020dsf,ATLAS:2020xzu,CMS:2021beq}, at least within
the framework of simplified models which are used for many LHC search
results.

In benchmark scenarios like the $m_h^{125}$ or the hMSSM\cite{Djouadi:2013uqa,Djouadi:2015jea}\footnote{In the hMSSM, the light Higgs mass is used as an input to ensure that $m_h =125$ GeV throughout the heavy Higgs search plane.},
it is hard to understand why the magnitude of the weak scale
$m_{weak}\sim m_{W,Z,h}$ is only $\sim 100$ GeV whilst sparticles,
especially higgsinos, are at the TeV or beyond scale.
This brings up the SUSY naturalness question\cite{Baer:2020kwz}
since it may be hard to maintain the MSSM as a plausible theory unless
it naturally accommodates the measured value of the weak scale.

In this work, we adopt the measured value of the $Z$-boson mass as
representative of the magnitude of weak scale,
where in the MSSM the $Z$ mass is related to the weak scale
Lagrangian parameters via the electroweak minimization condition
\be
m_Z^2/2 =\frac{m_{H_d}^2+\Sigma_d^d-(m_{H_u}^2+\Sigma_u^u )\tan^2\beta}{\tan^2\beta -1}-\mu^2
\label{eq:mzs}
\ee
where $m_{H_u}^2$ and $m_{H_d}^2$ are the Higgs soft breaking masses,
$\mu$ is the (SUSY preserving) superpotential $\mu$ parameter and
the $\Sigma_d^d$ and $\Sigma_u^u$ terms contain a large assortment of
loop corrections (see Appendices of Ref's \cite{Baer:2012cf} and \cite{Baer:2021tta}
and also \cite{Dedes:2002dy} for leading two-loop corrections).
Here, we adopt the notion of practical naturalness\cite{Baer:2015rja},
wherein an observable ${\cal O}$ is natural if all {\it independent}
contributions to ${\cal O}$ are comparable to\footnote{Here,
  the word {\it comparable} means to within a factor of a few.}
or less than ${\cal O}$.
For natural SUSY models, we use the naturalness measure\cite{Baer:2012up,Baer:2012cf}
\be
\Delta_{EW}\equiv |{\rm maximal\ term\ on\ the\ right-hand-side\ of\ Eq.~(\ref{eq:mzs})}|/(m_Z^2/2)\;,
\ee
where a value
\be
\Delta_{EW}\alt 30
\label{eq:dew30}
\ee
is adopted to fulfill the {\it comparable} condition of practical naturalness.
For most SUSY benchmark models, the superpotential $\mu$ parameter is tuned
to cancel against large contributions to the weak scale from SUSY breaking.
Since the $\mu$ parameter typically arises from very different physics
than SUSY breaking, {\it e.g.} from whatever solution to the SUSY
$\mu$ problem that is assumed,\footnote{Twenty solutions to the SUSY
  $\mu$ problem are recently reviewed in Ref. \cite{Bae:2019dgg}.}
then such a ``just-so'' cancellation seems highly implausible\cite{Baer:2022dfc}
(though not impossible) compared to the case where all contributions
to the weak scale are $\sim m_{weak}$,
so that $\mu$ (or any other parameter) need not be tuned.

There are several important implications of Eq. (\ref{eq:dew30}) for
heavy neutral SUSY Higgs searches.
\begin{itemize}
\item The superpotential $\mu$ parameter enters $\Delta_{EW}$ directly,
  leading to $|\mu |\alt 350$ GeV.
  This implies that for heavy Higgs searches with $m_{A,H}\agt 2|\mu |$, then
  SUSY decay modes of $H,\ A$ should typically be open. If these additional
  decay widths to SUSY particles are large, then the branching fractions to
  the (usually assumed) SM search modes may be substantially reduced.
\item For $m_{H_d}\gg m_{H_u}$, then $m_{H_d}$ sets the heavy Higgs mass scale
  ($m_{A,H}\sim m_{H_d}$) while $m_{H_u}$ sets the mass scale for $m_{W,Z,h}$.
  Then naturalness requires\cite{Bae:2014fsa}
    \end{itemize}
\be
m_{A,H}\alt m_Z\tan\beta\sqrt{\Delta_{EW}}.
\ee
For $\tan\beta\sim 10$ with $\Delta_{EW}\alt 30$, then $m_A$ can range up
to $\sim 5$ TeV. For $\tan\beta\sim 40$, then $m_A$ stays natural up to
$\sim 20$ TeV (although for large $\tan\beta\agt 20$, then bottom squark
contributions to $\Sigma_u^u$ become large and provide much
stronger upper limits on natural SUSY spectra\cite{Baer:2015rja}).

Since most heavy Higgs boson searches assume dominant $H,\ A\to SM$ decay modes,
then such results can overestimate the collider reach for these
particles. This is because,
in general, the presence of $H,\ A\to SUSY$ decay modes will diminish
heavy Higgs boson branching fractions to SM particles
via {\it e.g.} $H,\ A\to\tau\bar{\tau}$ or $b\bar{b}$ decays.

The most lucrative $H$ and $A$ search mode for $m_{H,A}\alt 1$ TeV
appears to be via the $H,\ A\to\tau\bar{\tau}$ mode.  This decay mode is
enhanced at large $\tan\beta$, and, unlike the $H,\ A\to b\bar{b}$ decay
mode, does not suffer from large QCD backgrounds.  Furthermore, the
narrow, low charge multiplicity jets that emerge from $\tau$ decay can
readily be identified. In Ref's \cite{ATLAS:2020zms,CMS:2022goy}, the
ATLAS and CMS collaborations used tau-jet identification along with the
cluster transverse mass variable to extract back-to-back (BtB) ditau
signal events in their heavy Higgs search results.  Results were derived
using $\sqrt{s}=13$ TeV $pp$ collisions with $\sim 139$ fb$^{-1}$ of
integrated luminosity.  No signal above background was seen, so limits
were placed in the $m_A$ vs. $\tan\beta$ plane assuming simplified heavy
Higgs benchmark scenarios such as
hMSSM\cite{Djouadi:2013uqa,Djouadi:2015jea} or $m_h^{125}$ scenario from
Ref. \cite{Bagnaschi:2018ofa}.  In these scenarios, the SUSY particles
are assumed too heavy to substantially influence the
$H,\ A\to\tau\bar{\tau}$ branching fractions.  Typical limits from
ATLAS\cite{ATLAS:2020zms} are that, for $\tan\beta =10$, then $m_A\agt
1.1$ TeV while for $\tan\beta =40$, then $m_A\agt 1.8$ TeV.  In
addition, in the same (or similar) scenarios, the projected HL-LHC reach
for $H,\ A\to\tau\bar{\tau}$ was estimated assuming $\sqrt{s}=14$ TeV
and 3000 fb$^{-1}$ of integrated
luminosity\cite{Cepeda:2019klc,Bahl:2020kwe}.  In these studies, the
95\% CL LHC reach for $H,\ A\to\tau\bar{\tau}$ for $\tan\beta =10$
extended out to $m_A\sim 1.35$ TeV and for $\tan\beta =40$ out to
$m_A\sim 2.25$ TeV.

In Ref. \cite{Baer:2022qqr}, we previously examined the LHC Run 3 and
HL-LHC reach for $H,\ A\to \tau\bar{\tau}$ in a natural SUSY benchmark
model, dubbed $m_h^{125}({\rm nat})$. In that benchmark, the lightest
electroweakinos (EWinos) are higgsino-like, with mass just a few hundred
GeV.  The, heaviest EWinos are wino-like, with mass $\sim 1$ TeV. Thus,
once $m_{H,A}\agt m(higgsino)+m(wino)$, then the decay modes $H,A\to
wino+higgsino$ (which proceed via the {\em unsuppressed}
gaugino-higgsino-Higgs boson coupling) become dominant unless $tan\beta$
is very large, at mass
values $m_{A,H}$ in the range of LHC search limits.  Using the perhaps
more plausible $m_h^{125}({\rm nat})$ scenario, the search limits become
reduced compared to LHC search results due to the turn-on of the
dominant supersymmetric decay modes.  In Ref. \cite{Baer:2022qqr}, also
a non-back-to-back ditau signal is also explored which allows for a
ditau invariant mass value $m_{\tau\tau}$ to be computed on an
event-by-event bases, with $m_{\tau\tau}$ yielding a (broad) peak around
$m_{\tau\tau}\sim m_{H,A}$.  The $m_{\tau\tau}$ distribution helps
separate signal from SM backgrounds, especially those arising from
$Z\to\tau\bar{\tau}$. By combining the BtB and non-BtB channels, the
discovery/exclusion reach is somewhat enhanced compared to using just
the BtB channel.

Taking heed that for $m_{A,H}\agt m(wino)+m(higgsino)$ then heavy Higgs
decays to EWino pairs becomes dominant, in the present paper we examine
prospects for LHC discovery/exclusion by looking at Higgs signals from
these supersymmetric decay modes.  Decays of heavy SUSY Higgs boson to
SUSY particles were originally explored in Ref's
\cite{Baer:1987eb,Gunion:1987ki,Gunion:1988yc}, but only in
Ref. \cite{Bae:2014fsa} were these decay modes examined in the context
of natural SUSY.  In that work, it was noted that for $H,\ A\to
wino+higgsino$ channels, the higgsino decays led to mainly soft,
quasi-visible decay debris whilst the winos decayed dominantly via
two-body modes into $W+higgsino$, $Z+higgsino$ and $h+higgsino$.  The
dominant search channels could then be categorized as 1. $h\to
b\bar{b}+\eslt$, 2. $Z\to\ell\bar{\ell}+\eslt$ and 3. $W\to\ell\nu_{\ell}
+\eslt$.  The last of these seemed plagued by huge backgrounds from SM
processes such as $W+jets$ and $WZ$ production while the first two also
appeared daunting.  Also, an $H,\ A\to 4\ell$ signature was examined in
Ref. \cite{Baer:2021qxa}, but rates appeared too low to be a viable
signature at LHC, although a signal at the Future Circular hadron-hadron
Collider, FCChh, operating with $\sqrt{s}=100$ TeV, appeared feasible
over some mass range.  In Ref's \cite{Barman:2016kgt} and
\cite{Adhikary:2020ujn}, some of these same signatures were also
examined, although mainly in the context of pMSSM instead of natural
SUSY.

In Sec. \ref{sec:decay} of this paper, we examine production cross
sections along with the branching ratios for the dominant decays of the
heavy Higgs $H$ and $A$ of the MSSM.  Over a wide range of parameters,
the SUSY modes dominate the SM decay modes once the kinematic decay
thresholds are passed.  In Sec. \ref{sec:channels}, we identify the main
final state channels which are available for discovery of $H,\ A\to
SUSY$ in natural SUSY models.  In Sec. \ref{ssec:WMET}, we examine the
$W(\to \ell\nu)+\eslt$ signal channel, and confirm that this is swamped by
SM background, at least at LHC luminosity upgrades.  In
Sec. \ref{ssec:ZMET}, we examine the $Z(\to\ell\bar{\ell})+\eslt$ channel,
in Sec. \ref{ssec:hMET}, we examine the $h(\to b\bar{b})+\eslt$ channel
and in Sec.\ref{ssec:1lLRjMET} we study the signal from $h$ or
$Z(\to\tau\bar{\tau})+\eslt$ events. We identify the $h(\to b\bar{b})+\eslt$
channel as the most promising.  We also examine whether the signal in
these channels can be further enhanced over SM backgrounds by requiring
additional soft leptons from the subsequent decays of the higgsinos. In
Sec. \ref{sec:results}, we combine signal significance from these
various channels (and others containing soft leptons coming from light
higgsino decays) to plot expected HL-LHC discovery and exclusion
contours in the $m_A$ vs. $\tan\beta$ plane.  Typically, these new
$H,\ A\to SUSY$ discovery channels may be accessible at HL-LHC for
$m_A\sim 1-2$ TeV (depending somewhat on $\tan\beta$).  Our summary and
conclusions are contained in Sec. \ref{sec:conclude}.

\section{Production of $H$ and $A$ followed by dominant decay to SUSY particles}
\label{sec:decay}

\subsection{A natural SUSY benchmark point}

For illustrative purposes, we here adopt a similar natural SUSY benchmark
point as in Ref. \cite{Baer:2022qqr}, which was dubbed $m_h^{125}({\rm
  nat})$ since the value of $m_h$ is very close to its measured value
throughout the entire $m_A$ vs. $\tan\beta$ plane.  We use the
two-extra-parameter non-universal Higgs model (NUHM2)\cite{Baer:2005bu}
with parameter space $m_0,\ m_{1/2},\ A_0,\ \tan\beta,\ \mu,\ m_A$ which
is convenient for naturalness studies since $\mu$ can be set to its
natural range of $\mu\sim 100-350$ GeV whilst both $m_A$ and $\tan\beta$
are free parameters.\footnote{The NUHM2 framework allows for independent
  soft SUSY breaking mass parameters for the scalar fields $H_u$ and
  $H_d$ in the Higgs sector, but leaves the matter scalar mass
  parameters universal to avoid flavour problems. The parameters
  $m_{H_u}^2$ and $m_{H_d}^2$ are then traded for $\mu$ and $m_A$ in
  Eq. (\ref{eq:param}).} We adopt the following natural SUSY
benchmark Higgs search scenario: \be m_h^{125}({\rm nat}):\ m_0=5\ {\rm
  TeV},\ m_{1/2}=1\ {\rm TeV},\ A_0=-1.6m_0,\ \tan\beta
,\ \mu=200\ {\rm GeV}\ {\rm and}\ m_A . \label{eq:param} \ee 

A similar $m_h^{125}({\rm nat})$ benchmark model spectrum, but with
$\mu =250$ GeV and $m_{1/2}=1.2$ TeV, was shown in
Table 1 of Ref. \cite{Baer:2022qqr} for $\tan\beta =10$ and $m_A=2$ TeV
and so for brevity we do not show the revised spectrum here.  
We adopt the computer code
Isajet\cite{Paige:2003mg} featuring Isasugra\cite{Baer:1994nc} for
spectrum generation and computation of dark matter observables and other
low energy observables for comparison with data.\footnote{We note that
  the value of $\sigma^{SI}(\tchi_1^0,p)$ from Table 1 of
  Ref. \cite{Baer:2022qqr} appears to be in conflict with the recent
  bounds from the LZ experiment\cite{LZ:2022ufs} by about a factor of 3
  (see also results from Xenon1T\cite{XENON:2018voc} and
  PandaX-II\cite{PandaX-II:2017hlx}) on direct detection of WIMP
  scattering on their liquid Xe target, even taking into account that
  the relic neutralinos are thermally underproduced with the remainder
  of dark matter composed of {\it e.g.}  axions\cite{Baer:2016ucr} or
  something else.  We are not particularly concerned by this, since it
  is easy to imagine that entropy dilution from late decaying saxions
  \cite{Bae:2014rfa} or moduli\cite{Bae:2022okh} fields could further
  reduce the neutralino relic abundance bringing the BM point into
  accord with limits from direct detection\cite{Gelmini:2006pw} with no
  impact upon the LHC phenomenology discussed in this paper.
  A similar situation obtains for the benchmark case
    considered here.}  The SUSY
Higgs boson masses are computed using renormalization-group (RG)
improved third generation fermion/sfermion loop
corrections\cite{Bisset:1995dc}.  The RG improved Yukawa couplings
include full threshold corrections\cite{Pierce:1996zz} which account for
leading two-loop effects\cite{Carena:2002es}.  For $\tan\beta =10$ and
$m_A=2$ TeV, we note that $\Delta_{EW}=16$ so the model is indeed EW
natural.  Also, with $m_h=124.6$ GeV, $m_{\tg}=2.4$ TeV and
$m_{\tst_1}=1.6$ TeV, it is consistent with LHC Run 2 SUSY search
constraints.  Most relevant to this paper, the two lightest neutralinos,
$\tchi_1^0$ and $\tchi_2^0$, and the lighter chargino, $\tchi_1^\pm$,
are higgsino-like with masses $\sim 200$ GeV while the neutralino
$\tchi_3^0$ is bino-like with a mass of 450~GeV and the heaviest
neutralino and the heavier chargino have masses $\sim 0.86$~TeV.
Thus, the $H,\ A\to wino+higgsino$ decay modes turn on for
$m_{H,A}\agt 1.1$ TeV (although $H,\ A\to bino+higgsino$ turns on at
somewhat lower $m_A$ values).

\subsection{$H$ and $A$ production cross sections at LHC}
\label{ssec:sigma}

\begin{figure}[htb!]
\begin{center}
\includegraphics[height=0.25\textheight]{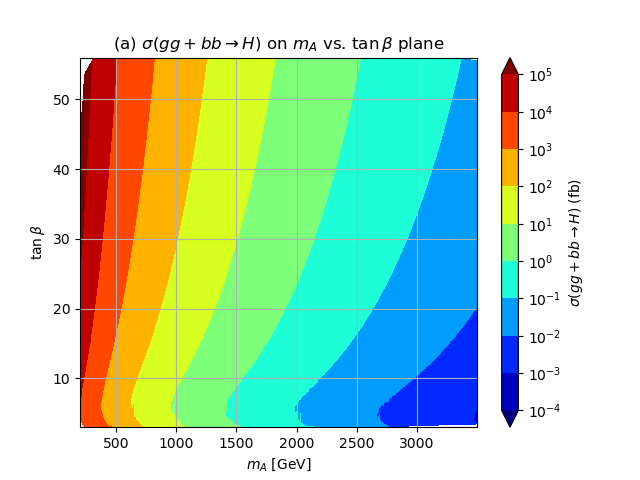}
\includegraphics[height=0.25\textheight]{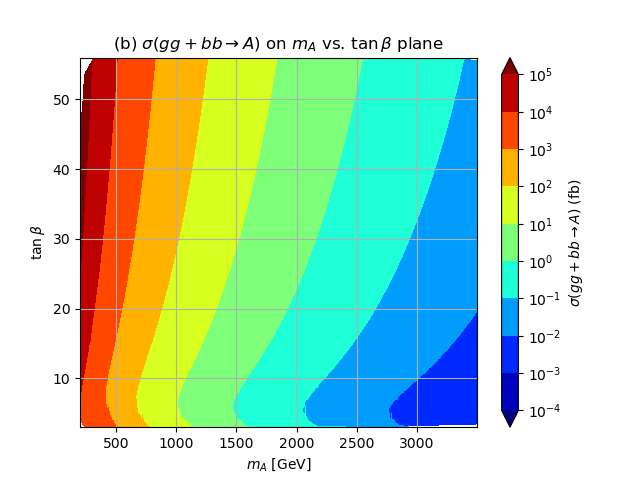}\\
\caption{The total cross section for {\it a}) $pp\to H$
  and {\it b}) $pp\to A$ at $\sqrt{s}=14$ TeV using the SusHi
  code\cite{Harlander:2012pb}.
\label{fig:sigHA14}}
\end{center}
\end{figure}

In Fig. \ref{fig:sigHA14}, we show total cross sections for {\it a})
$pp\to H+X$ and {\it b}) $pp\to A+X$ production in the $m_A$ vs. $\tan\beta$
plane using the computer code SusHi\cite{Harlander:2012pb} which
includes leading NNLO corrections.  The plots show $\sigma$ in fb units
for $\sqrt{s}=14$ TeV.  The dominant production processes come from the
$gg$ and $b\bar{b}$ fusion diagrams, with the latter dominating unless
$\tan\beta$ is very small. In the Figure, the total production cross
section is color coded with cross sections ranging as high as
$\sigma\sim 10^5$ fb on the left edge, although this region is now
LHC-excluded.  For the LHC-allowed regions where {\it e.g.} $m_A\agt 1$
TeV for $\tan\beta\sim 10$, then the cross sections lie typically below
10 fb.  Of course, the $\sigma$ values drop off for increasing $m_A$ but
also we see how they increase for increasing $\tan\beta$.  For any given
value of $m_A$ and $\tan\beta$, the production cross sections for $pp\to
H$ and $pp\to A$ are typically very close in value to each other.

\subsection{$H$ and $A$ branching fractions}
\label{ssec:BF}

\begin{figure}[htb!]
\begin{center}
\includegraphics[height=0.2\textheight]{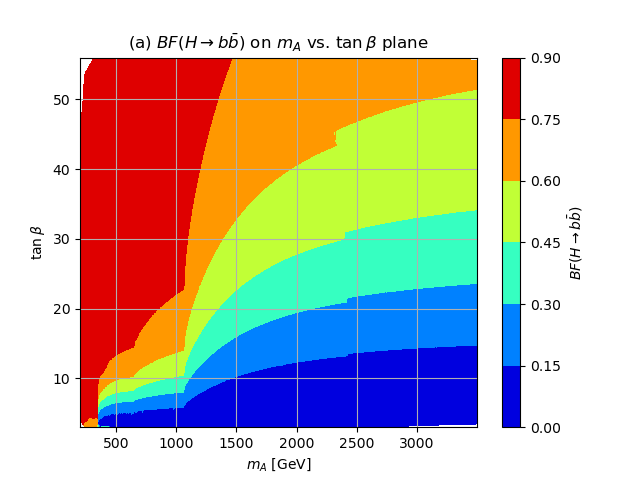}
\includegraphics[height=0.2\textheight]{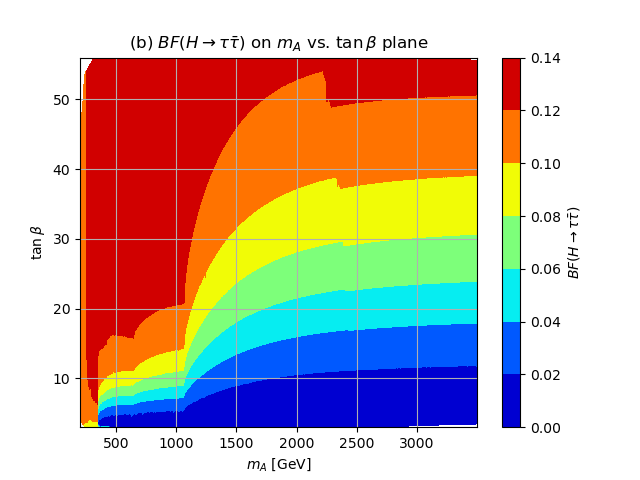}\\
\includegraphics[height=0.2\textheight]{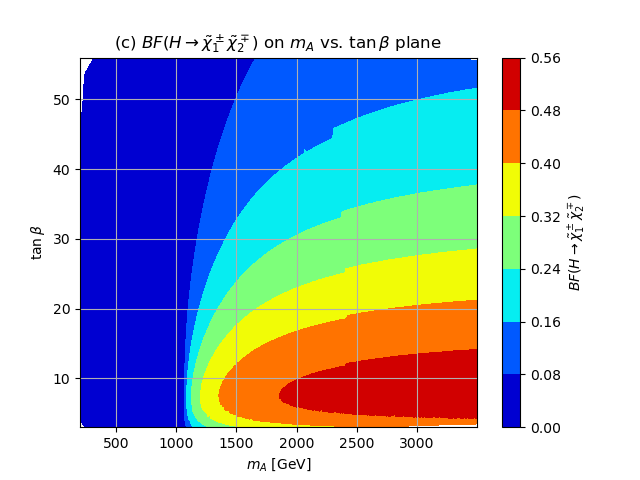}
\includegraphics[height=0.2\textheight]{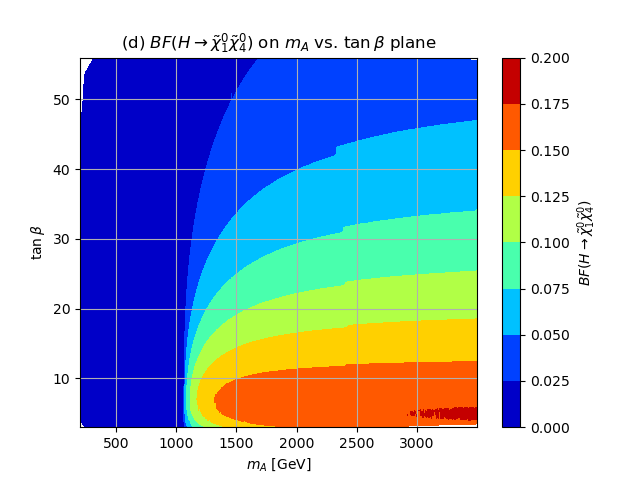}\\
\includegraphics[height=0.2\textheight]{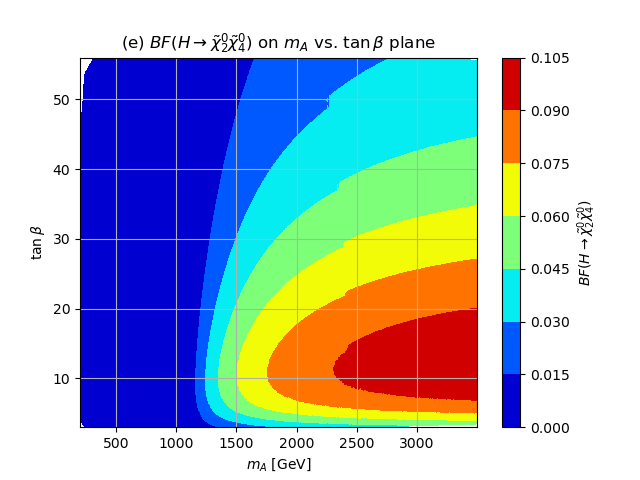}
\includegraphics[height=0.2\textheight]{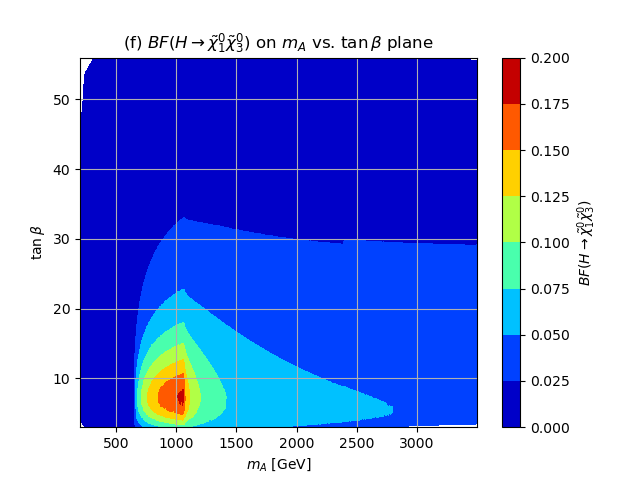}\\
\caption{Branching fractions for $H$ to {\it a}) $b\bar{b}$,
  {\it b}) $\tau\bar{\tau}$, {\it c}) $\tchi_1^\pm\tchi_2^\mp$,
  {\it d}) $\tchi_1^0\tchi_4^0$, {\t e}) $\tchi_2^0\tchi_4^0$
  and {\it f}) $\tchi_1^0\tchi_3^0$ from Isajet 7.88\cite{Paige:2003mg}.
\label{fig:BFH}}
\end{center}
\end{figure}

\begin{figure}[htb!]
\begin{center}
\includegraphics[height=0.2\textheight]{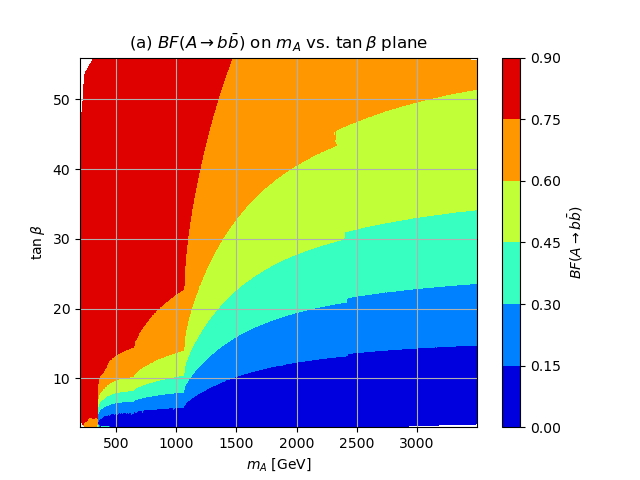}
\includegraphics[height=0.2\textheight]{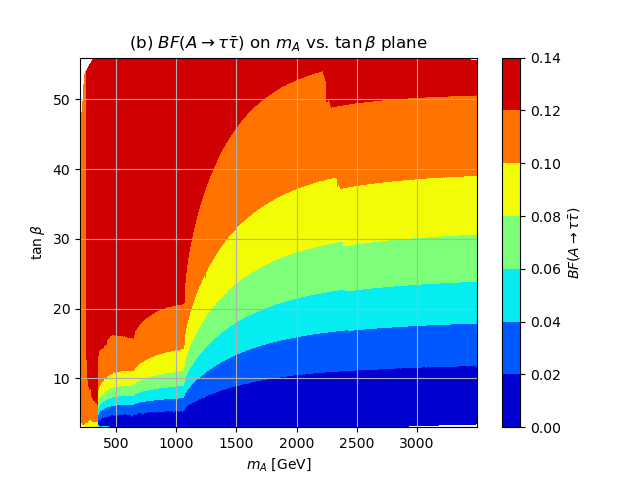}\\
\includegraphics[height=0.2\textheight]{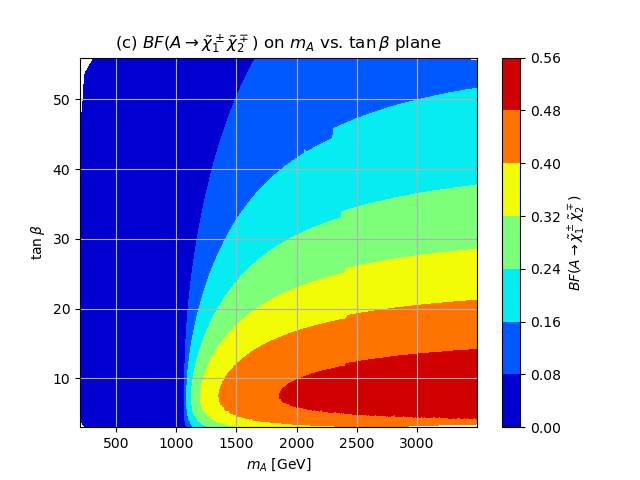}
\includegraphics[height=0.2\textheight]{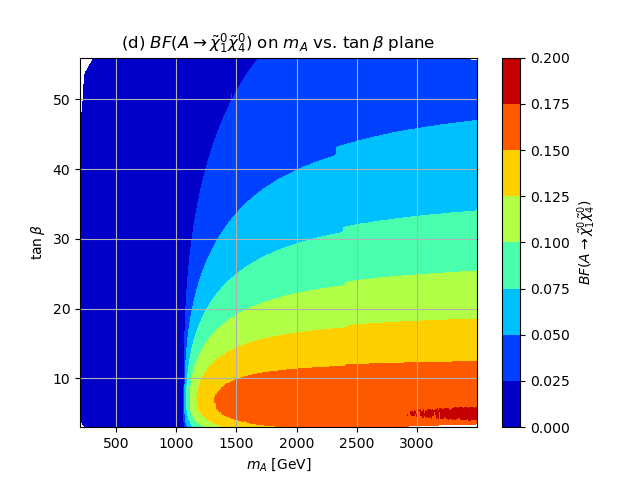}\\
\includegraphics[height=0.2\textheight]{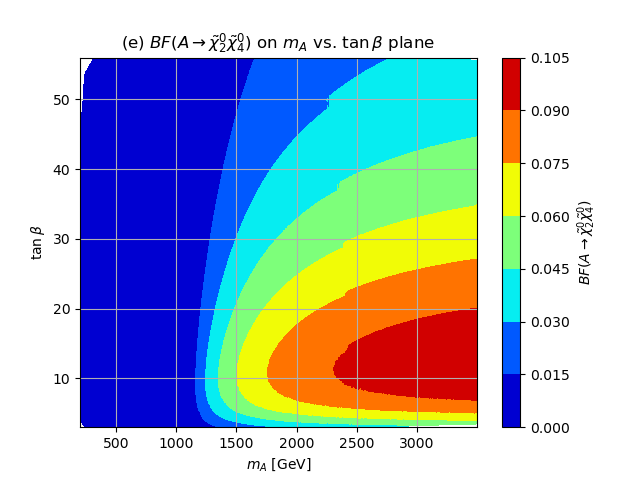}
\includegraphics[height=0.2\textheight]{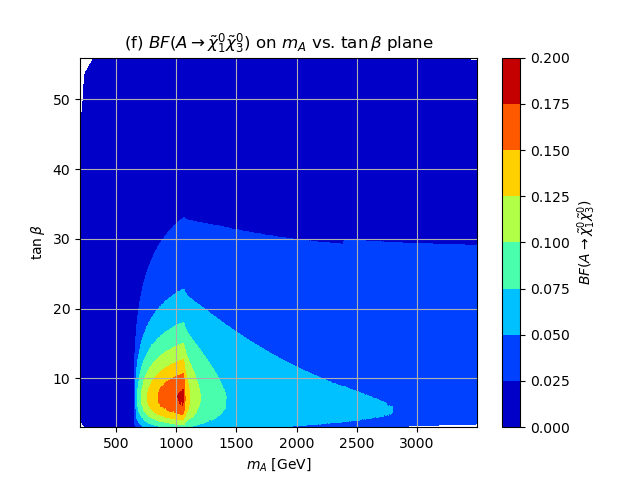}\\
\caption{Branching fractions for $A$ to {\it a}) $b\bar{b}$,
  {\it b}) $\tau\bar{\tau}$, {\it c}) $\tchi_1^\pm\tchi_2^\mp$,
  {\it d}) $\tchi_1^0\tchi_4^0$, {\t e}) $\tchi_2^0\tchi_4^0$
  and {\it f}) $\tchi_1^0\tchi_3^0$ from Isajet 7.88\cite{Paige:2003mg}.
\label{fig:BFA}}
\end{center}
\end{figure}

In Figures \ref{fig:BFH} and \ref{fig:BFA}, we show some select
$H$ and $A$ branching fractions (BFs) in the $m_A$ vs. $\tan\beta$ plane.
The branching fractions are again color-coded, with the larger ones
denoted by red whilst the smallest ones are denoted by dark blue.
The branching fractions are extracted from the Isasugra code\cite{Paige:2003mg}
and the decay formulae can be found in Appendix C of Ref. \cite{Baer:2006rs}.

In Fig. \ref{fig:BFH}{\it a}) we show the BF for $H\to b\bar{b}$.
This decay mode to SM particles is indeed dominant for $m_A\alt 1$ TeV and
for larger values of $\tan\beta\agt 10-20$.
In frame {\it b}), we show the BF($H\to\tau\bar{\tau}$).
Like $H\to b\bar{b}$, this mode is enhanced at large $\tan\beta$ and has
provided the best avenue for heavy Higgs discovery/exclusion plots so far.

While SUSY decay modes of $H$ and $A$ to higgsino pairs are also open in
these regions, these decay modes is suppressed by mixing angles.  In the
MSSM, there is a direct gauge coupling\cite{Baer:2006rs} \be {\cal
  L}\ni-\sqrt{2}\sum_{i,A}{\cal S}_i^\dagger g t_A\bar{\lambda}_A\psi_i
+H.c.  \ee where ${\cal S}_i$ labels various matter and Higgs scalar
fields, $\psi_i$ is the fermionic superpartner of ${\cal S}_i$ and
$\lambda_A$ is the gaugino with gauge index $A$.  Also, $g$ is the
corresponding gauge coupling for the gauge group in question and the
$t_A$ are the corresponding gauge group matrices.  Letting ${\cal S}_i$
be the Higgs scalar fields, we see there is an unsuppressed coupling of
the Higgs scalars to gaugino plus higgsino as mentioned earlier.  This
coupling can lead to dominant SUSY Higgs boson decays to SUSY particles
when the gaugino-plus-higgsino decay channel is kinematically
unsuppressed. But it also shows why the heavy Higgs decay to higgsino
pairs is suppressed by mixing angles for $|\mu| \ll |M_{1,2}|$, once
we recognize that a Higgs boson-higgsino-higgsino coupling is forbidden
by gauge invariance. 

In frame {\it c}), we show BF($H\to \tchi_1^\pm\tchi_2^\mp$), where
$\tchi_1^\pm$ is dominantly higgsino-like and $\tchi_2^\pm$ is
dominantly wino-like for natural SUSY models like the $m_h^{125}({\rm nat})$
scenario. Here, we see that for larger values of $m_A\simeq m_H\agt 1.2$ TeV,
then this mode turns on, and at least for moderate $\tan\beta\sim 10-20$
(which is favored by naturalness\cite{Bae:2014fsa}), rapidly comes to dominate
the $H$ decay modes along with the neutral wino+higgsino channels
$H\to \tchi_1^0\tchi_4^0$ (frame {\it d})) and $H\to \tchi_2^0\tchi_4^0$
(frame {\it e})). Here, $\tchi_4^0$ is mainly neutral wino-like
while $\tchi_{1,2}^0$ are mainly higgsino-like. The sum of these
three wino+higgsino decay channels thus dominate the $H$ decay branching
fractions for $m_{H,A}\agt 1.2$ TeV and low-to-moderate values of
$\tan\beta$. For high values of $\tan\beta$, the bottom and $\tau$ Yukawa
couplings become large, and SM decays to fermions once again dominate
SUSY decays. Decays of $H$ to gauge boson pairs are unimportant
in the decoupling limit.
For completeness, we also show in frame {\it f}) the decay mode
$H\to\tchi_1^0\tchi_3^0$ which is to higgsino+bino. This mode is large
only  in a
small region of $m_H\sim 1$ TeV and modest $\tan\beta$ where the mode
$H\to bino+higgsino$ decay has turned on,
but where $H\to wino+higgsino$ has yet to become kinematically
open. Decays to winos dominate decays to binos because
the $SU(2)$ gauge coupling is larger than the hypercharge gauge coupling.

In Fig. \ref{fig:BFA}, we show the same branching fractions as in
Fig. \ref{fig:BFH}, but this time for $A$ decay.
The plots are very similar to the results from Fig. \ref{fig:BFH},
and for largely the same reasons. For $m_A\agt 1.2$ TeV and
small-to-moderate $\tan\beta$, then $A\to$ wino+higgsino becomes the
dominant $A$ decay mode. We note that $A$ does not couple to vector
boson pairs. 

\section{A survey of various $H,\ A\to SUSY$ signal channels}
\label{sec:channels}

Having established that the SUSY decay modes of $H$ and $A$ may
dominate soon after they become kinematically allowed, we explore the
ensuing $H,\ A\to SUSY$ signatures for LHC upgrades in order to
determine if they can open new avenues to discovery or, perhaps,
confirmation of a signal seen via a SM channel. The dominant $H$ and $A$
decay modes are to neutral and charged winos plus light higgsinos where
the light higgsinos, if they are not LSP, decay to very soft visible SM
particles.  The $\tchi_4^0$, which in our case is mainly neutral wino,
typically decays via $\tchi_4^0\to W^\pm\tchi_{1}^\mp$ with a branching
fraction $\sim 50\%$, while the decays $\tchi_4^0\to Z\tchi_{1,2}^0$
and to $\tchi_4^0\to h\tchi_{1,2}^0$ each have branching fractions $\sim
25\%$. The $\tchi_2^\pm$, which is mainly charged wino, decays via
$\tchi_2^-\to W^-\tchi_{1,2}^0$ $\sim 50\%$ of the time, with
$\tchi_2^-\to Z\tchi_1^-$ and $h\tchi_1^-$ each have branching fractions
$\sim 25\%$.  Combining the decay patterns, the dominant
$H,\ A\to SUSY$ decay modes lead to the following signatures:
\bi
\item $H,\ A\to W+\eslt$,
\item $H,\ A\to Z+\eslt$ and
  \item $H,\ A\to h+\eslt$.
\ei
Each of these signatures may also contain some soft leptons or jets
which arise from the light higgsino decays; these soft visibles potentially
can lead to further discovery channels, or at least enhance the
discovery/exclusion channels if the SM backgrounds are under control.

For our simulations, we generate signal SUSY Les Houches Accord
(SLHA) output files using Isajet\cite{Paige:2003mg} and feed these into
Pythia\cite{Sjostrand:2007gs} for event generation.
We then interface Pythia with the Delphes\cite{deFavereau:2013fsa}
toy detector simulation code. The SM backgrounds of $pp\to W$,
$\gamma^*/Z$, $t\bar{t}$, $VV$ ($V=W$ or $Z$), $Wh$ and $Zh$ production
are all generated using Pythia.

For all events, we require they pass one of our baseline trigger
requirements:
\bi
\item Baseline small-radius jet (SRj): using anti-$k_T$ jet finder
  algorithm, require $p_T(SRj)>25$ GeV with jet cone size $\Delta R<0.4$
  and $|\eta_{SRj}|<4.5$.

\item Baseline large-radius jet (LRj): using anti-$k_T$ jet finder
  algorithm, require $p_T(LRj)>100$ GeV with jet cone size $\Delta
  R<1.2$ and $|\eta_{LRj}|<4.5$, for all signatures except for \newline 
  $1LRj+\ell +\eslt$, for which $\Delta R<1.5$. The large-radius
  jets are formed using calorimeter deposits (or track information for
  muons) so that even isolated leptons (see below) are included as
  constituents of these jets. This will be especially important for the signal
  examined in Sec.~\ref{ssec:1lLRjMET} below.   

\item Baseline isolated lepton: satisfy basic Delphes lepton isolation
  requirement with $p_T(\ell )>5$ GeV, lepton cone size $\Delta R<0.3$,
  and $pTRatio(e) < 0.1$ while $pTRatio(\mu) < 0.2$, where $pTRatio$ is defined
  in Delphes as ${\sum_i{|p_{Ti}|}}\over {p_T(\ell)}$, for calorimetric
    cells within the lepton cone.
\ei

For the signal search, we further require \bi
\item SRj: satisfy above SRj requirement plus $|\eta_{SRj}|<2.4$ .
\item $b$-jets: satisfy above SRj requirement plus $b$-jet tagged by Delphes
  $b$-tagging requirement.
\item signal leptons: require above baseline lepton qualities plus
  $p_T(e)>20$ GeV with $|\eta (e)|< 2.47$ while
  $p_T(\mu)>25$ GeV with $|\eta (\mu )|<2.5$.
\ei

We have examined several distributions for four cases with $m_A=1.5$ and
2~TeV, and $\tan\beta$=10 and 40 to arrive at suitable cuts for the
various signals from $H,A \to gaugino + higgsino$ decays that we discuss
in the remainder of this section.

\subsection{$H,\ A\to W(\to\ell\nu)+\eslt$ signal}
\label{ssec:WMET}

For the $H,\ A\to W+\eslt$ channel, we will look for $W\to \ell\nu_\ell$
where $\ell = e$ or $\mu$.  
\bi
\item exactly one baseline lepton
(and no LRjs which will comprise an alternative channel: see below).
This lepton should also satisfy signal lepton requirements.
\ei
  After examining various distributions, we require
\bi
\item $|\eta (\ell )|<1.3$,
\item $\eslt >150$ GeV,
\item $\Delta\phi (\ell, \vec{\eslt})>90^\circ$,
\item transverse mass $m_T(\ell ,\eslt )>100$ GeV and
  \item $p_T(W)>20$ GeV where $\vec{p}_T(W)=\vec{p}_T(\ell )+\vec{\eslt}$.
\ei

\begin{figure}[htb!]
\begin{center}
\includegraphics[height=0.4\textheight]{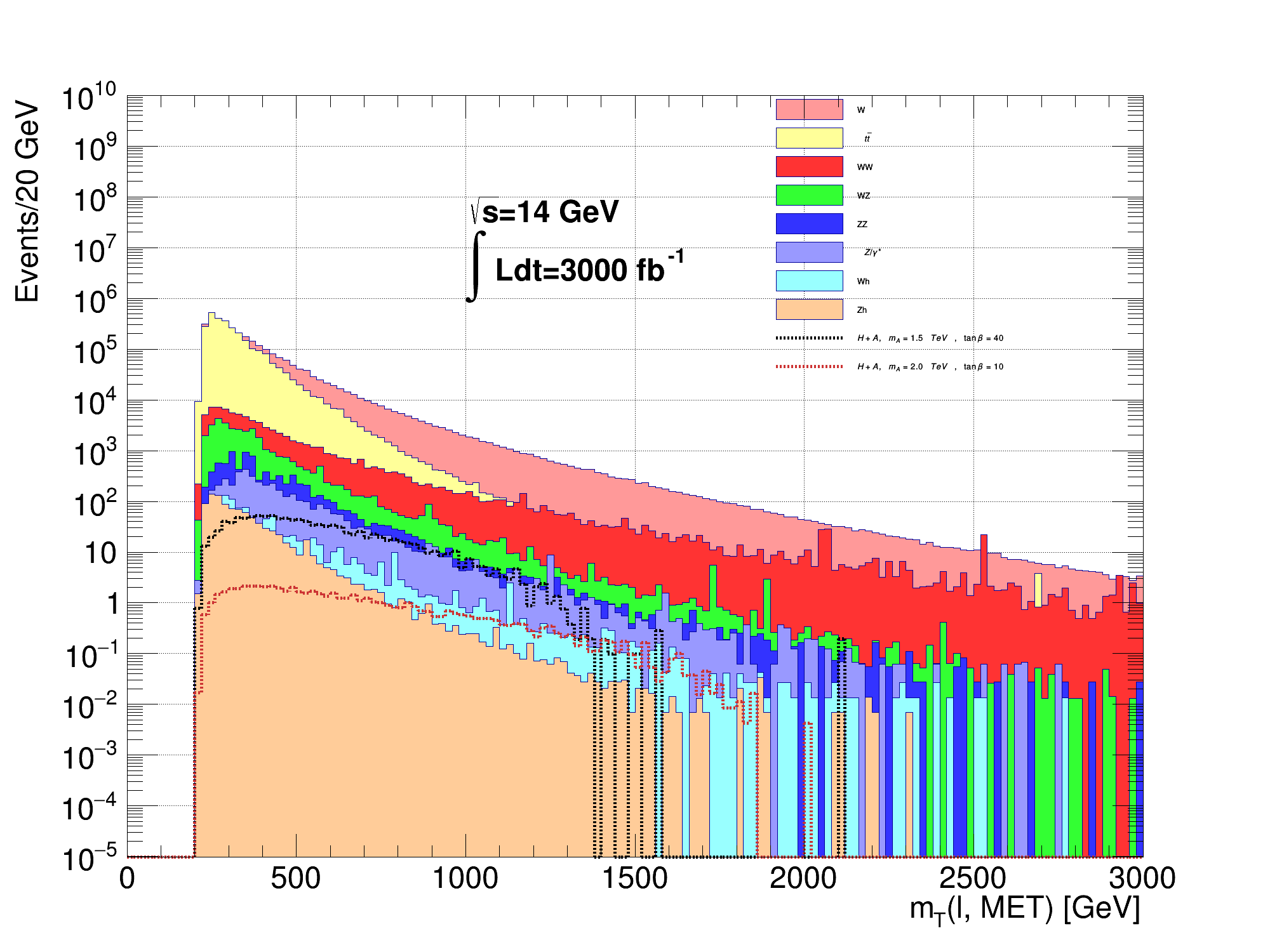}
\caption{Distribution in $m_T(\ell,\eslt )$ for $pp\to H,A\to W+\eslt$
  events. We show two signal distributions (dashed) along with
  dominant SM backgrounds (not stacked).
\label{fig:mTlMET}}
\end{center}
\end{figure}

Our goal in each signal channel is to look for an excess above the SM
backgrounds in the largest transverse mass bins which are most sensitive to
the TeV-scale heavy Higgs decay.  Our basic results are shown in
Fig. \ref{fig:mTlMET}, where we plot two signal benchmark cases:
one for $m_A=1.5$ with $\tan\beta =40$ (black-dashed) 
and one for $m_A=2$ TeV with $\tan\beta =10$ (orange-dashed), 
while the SM BG distributions are color coded as solid
(unstacked) histograms.\footnote{We are aware that the $m_A=1.5$~TeV,
  $\tan\beta=40$ BM case is just excluded at 95\%CL by the Atlas search for
  heavy Higgs bosons, assuming that $H,A$ decays essentially only via SM modes
  \cite{ATLAS:2020zms}.}  As anticipated\cite{Bae:2014fsa}, there is an
enormous SM BG from direct off-shell $W$ production, as indicated by the
pink histogram.  The next largest SM BGs come from $t\bar{t}$, $WW$ and
$WZ$ production (yellow, red and green histograms).  Over the entire
range of $m_T$, the SM BGs lie several orders of magnitude above our
SUSY BM models. Thus, it appears to be extremely difficult to root out a
signal via the single lepton channel.

\subsection{$H,\ A\to Z(\to \ell\bar{\ell})+\eslt$ signal}
\label{ssec:ZMET}

In this channel, we search for a high momentum, leptonically decaying $Z$
boson recoiling against $\eslt$.
Here, we require the following.
\bi
\item Exactly two baseline leptons (veto aditional leptons).
\item The two leptons satisfy signal lepton requirements and are
  opposite-sign/same flavor (OS/SF).
\item The dilepton pair invariant mass reconstructs $m_Z$:
  $80\ {\rm GeV}<m(\ell\bar{\ell})<100$ GeV.
\ei

We also require,
\bi
\item $\eslt >250$ GeV,
\item $\eslt_{,rel}:=$ $\eslt\cdot\sin{(min(\Delta\phi,\frac{\pi}{2}))}>125$ GeV, where $\Delta\phi$ is the azimuthal angle between the $\Vec{\eslt}$ and the closest lepton or jet with $p_T > 25$ GeV.
\item $|\eta (\ell_1)|<1.3$, $|\eta (\ell_2 )|<2$ ( $p_T(\ell_1) >
  p_T(\ell_2)$) and
  $|\eta (\ell\bar{\ell})|<1.5$, 
\item $\Delta\phi (\ell\bar{\ell},\eslt )>140^\circ$,
\item $\Delta\phi (\ell_1,\ell_2)<80^\circ$ and
  \item $|\vec{p}_T(Z)+\vec{\eslt} |>25$ GeV.
\ei

\begin{figure}[htb!]
\begin{center}
\includegraphics[height=0.4\textheight]{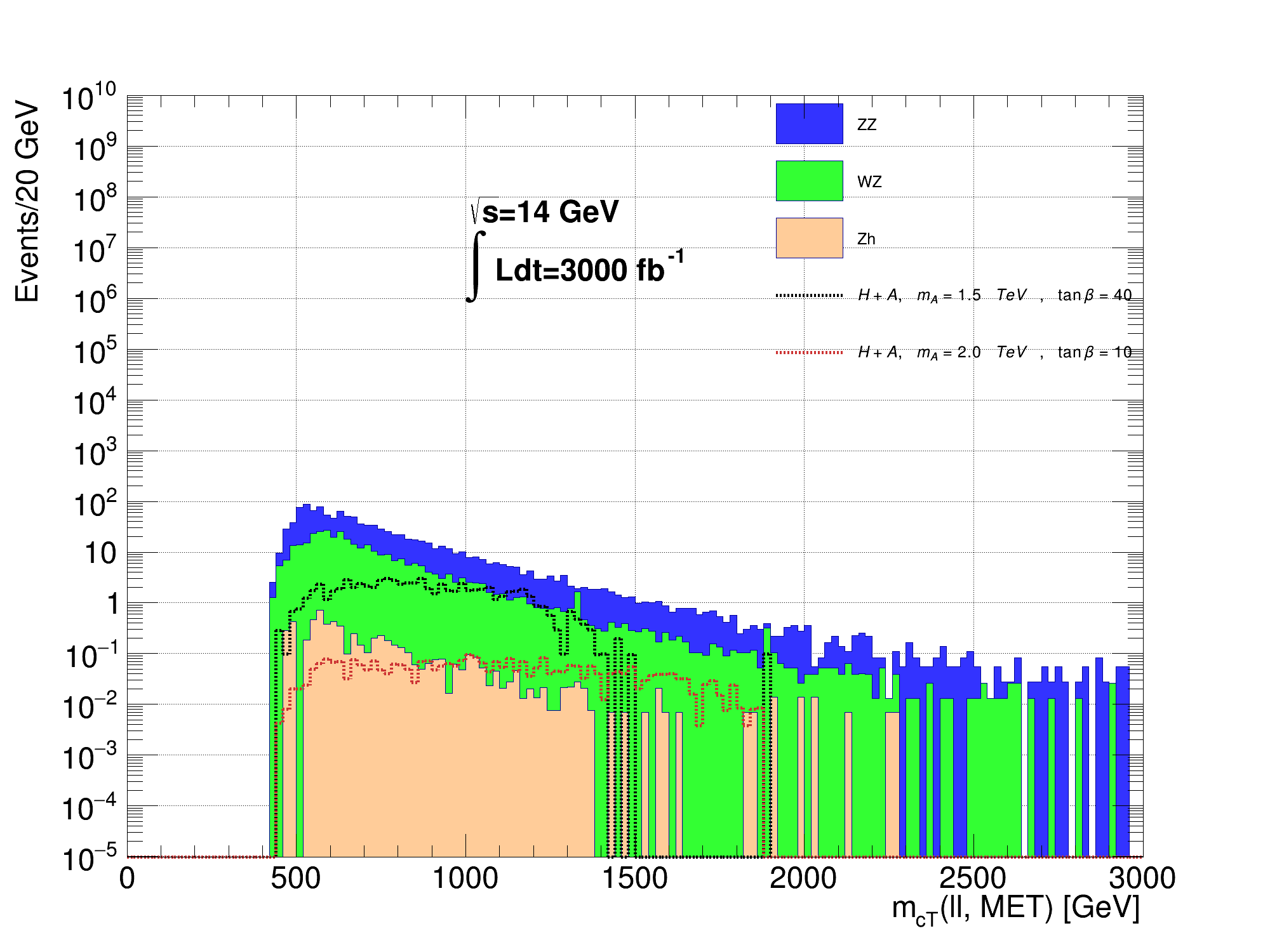}
\caption{Distribution in $m_{cT}(\ell\bar{\ell},\eslt )$ for
  $pp\to H,A\to Z+\eslt$ with $Z\to\ell^+\ell^-$ decay events
  at $\sqrt{s}=14$ TeV.
  We show two signal distributions (dashed) along with
  dominant SM backgrounds (not stacked).
\label{fig:mT2lMET}}
\end{center}
\end{figure}

We then plot the cluster transverse mass\cite{Barger:1984sm}
$m_{cT}(\ell\bar{\ell},\eslt )$ in Fig. \ref{fig:mT2lMET}.  From the
figure, we see that the dominant SM backgrounds come from the $ZZ$ and
$WZ$ production followed by subdominant $Zh$ production. For the signal
from our benchmark point with $m_A=1.5$ TeV and $\tan\beta =40$ (black-dashed histogram), we see that signal exceeds $WZ$ background around
$m_{cT}\sim 1200$ GeV and is only a factor of $\sim 2$ below the dominant
$ZZ$ BG.  Thus, we might expect a significant shape deviation in the
$m_{cT}(\ell\bar{\ell},\eslt)$ distribution at large transverse mass
values, signaling the presence of a heavy, new physics object
contributing to this distribution.

\subsection{$H,\ A\to h(\to b\bar{b})+\eslt$ signal}
\label{ssec:hMET}

In this case, we search for production of a single SM-like Higgs boson
recoiling against large $\eslt$, with $h\to b\bar{b}$.  
Here, we require
\bi
\item At least two tagged $b$-jets,
\item veto any baseline leptons,
\item exactly one LRj,
\item at least two $b$-jets within the cone of the LRj,
\item exactly one di-$b$-jet pair within the cone of the LRj reconstructs the light Higgs mass: $90\ {\rm GeV}<m(bb)<135$ GeV,
\item $m(LRj)-m(bb)>-5$ GeV
    \ei

    We further require:
\bi
\item $\eslt >225$ GeV,
\item $\eslt_{, rel}>225$ GeV,
\item $H_T>350$ GeV,
\item $p_T(b_1)>100$ GeV,
\item $p_T(bb)/p_T(LRj)>0.9$,
\item $m(bb)/p_T(LRj)>0.9$,
\item $|\eta (bb)|<1.4$,
\item $\Delta\phi(bb,\eslt )>145^\circ$,
\item  $max[\Delta\phi(j,LRj)]<140^\circ$ where $j$ loops over all baseline SR jets in the event and
\item $\Delta\phi (b_1,b_2)<65^\circ$.
\ei

\begin{figure}[htb!]
\begin{center}
\includegraphics[height=0.4\textheight]{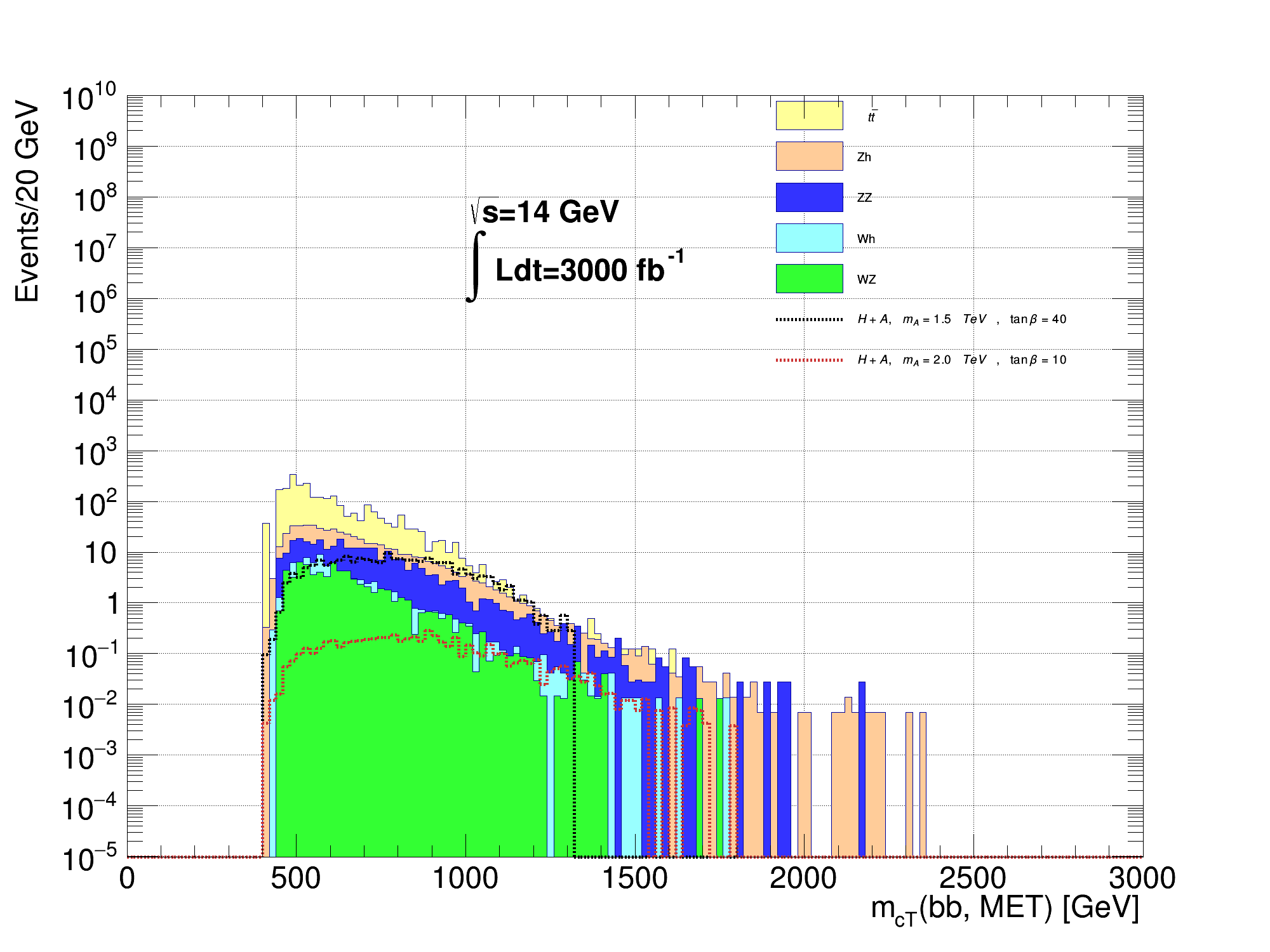}
\caption{Distribution in $m_{cT}(b\bar{b},\eslt )$ for
  $pp\to H,A\to h+\eslt$ with $h\to b\bar{b}$ decay events
  at $\sqrt{s}=14$ TeV.
  We show two signal distributions (dashed) along with
  dominant SM backgrounds (not stacked).
\label{fig:mT2bMET}}
\end{center}
\end{figure}

Here, $H_T$ is the scalar sum of $E_T$s of all visible objects
in the event\cite{Baer:1988kq}.
We plot the ensuing $m_{cT}(b\bar{b},\eslt )$ distribution in
Fig. \ref{fig:mT2bMET}. From the figure, we see that the dominant SM
backgrounds come from $t\bar{t}$ production (where the $b$-jets
accidently reconstruct the Higgs boson mass) followed by $Zh$ and then $ZZ$
production. In this case, at large $m_{cT}\agt 1$ TeV, the $\tan\beta =40$,
$m_A=1.5$ TeV BM case is actually comparable to SM backgrounds.
Thus, we would expect a measureable shape deviation in this
distribution at high values of cluster transverse mass, for at least
some of our BM points.

\subsection{$H,\ A\to 1LRj+ \ell +\eslt$ signal}
\label{ssec:1lLRjMET}

Here, we examine the prospects for observing the signal from $H, A$
decays that yield a high $p_T$ $Z$ or $h$ boson plus $\eslt$, where the
$Z$ or the $h$ decays into tau pairs and one of the taus decays
hadronically and the other leptonically. Such a topology will yield an
SRj plus an isolated signal lepton
that coalesce to a single LRj that includes an identified
lepton within a cone with $\Delta R < 1.5$.


For this channel, we require:
\bi
\item Exactly one signal lepton and no additional baseline leptons in the event.
\item Exactly one LRj with invariant mass $40\ {\rm GeV}<m(LRj)<145$ GeV
  in accord with a source of either $W$, $Z$ (or $h$).
\item The lepton is within the cone of the LRj.
\ei

In addition, we require
    \bi
  \item $\eslt >275$ GeV,
  \item $\eslt_{,rel}>125$ GeV,
  \item $|\eta (\ell )|<1.7$,
  \item $H_T>350$ GeV,
  \item $\Delta\phi(LRj,\eslt)>140^\circ$,
  \item $p_T(\ell )/p_T(LRj)>0.9$,
  \item no $b$-jet within cone of the LRj and
    \item no jets with $p_T(j)>100$ GeV outside the cone of the LRj.
    \ei

\begin{figure}[htb!]
\begin{center}
\includegraphics[height=0.4\textheight]{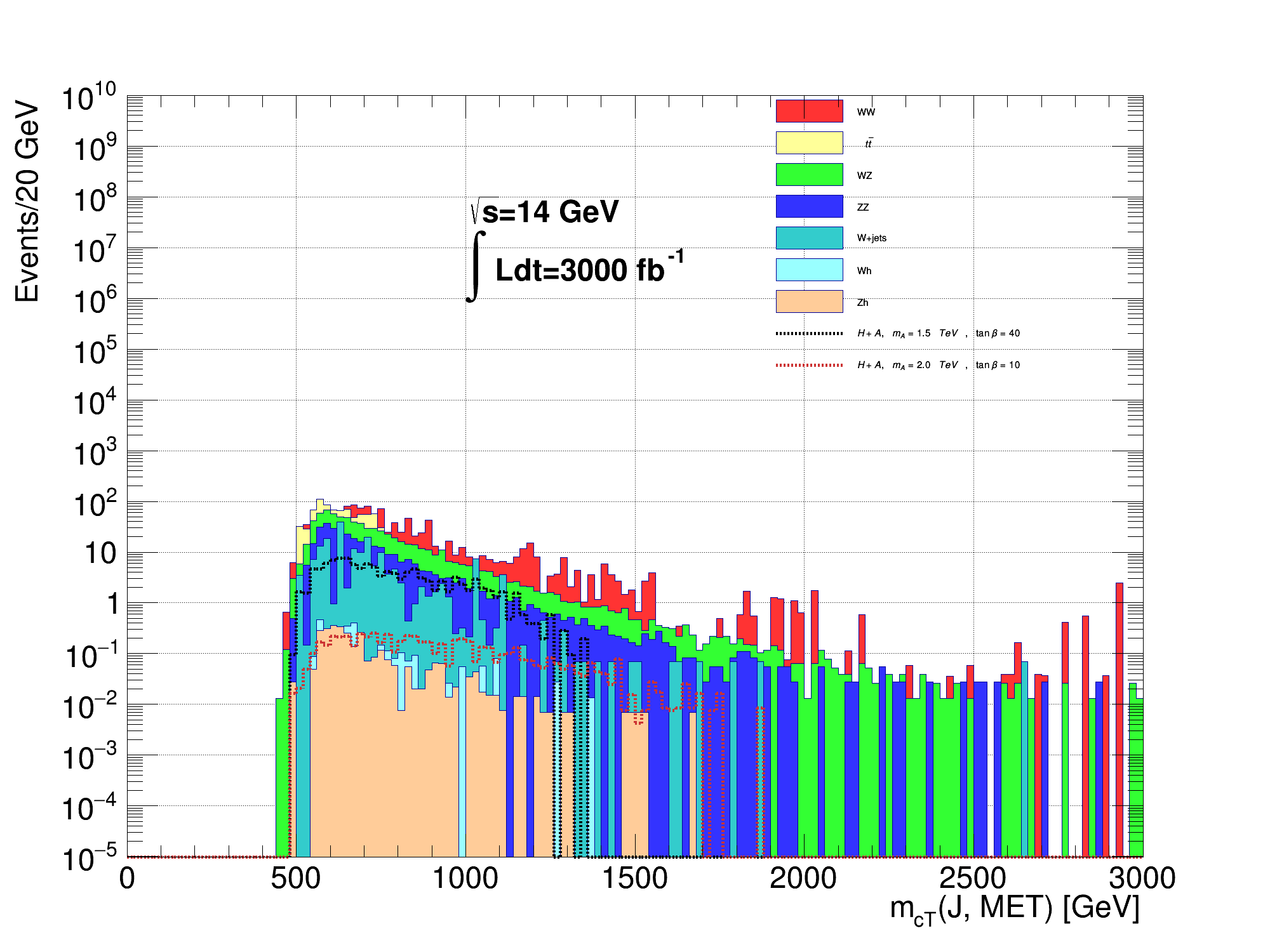}
\caption{Distribution in $m_{cT}(LRj, \eslt )$ for
  $pp\to H,A\to LRj+ \ell +\eslt$ events
  at $\sqrt{s}=14$ TeV.
  We show two signal distributions (dashed) along with
  dominant SM backgrounds (not stacked).
\label{fig:1lLRj}}
\end{center}
\end{figure}

    We next plot the resulting $m_{cT}(LRj,\eslt )$ distribution (not
    including the lepton) in Fig. \ref{fig:1lLRj}. From the plot, we see
    that the largest BGs come from $WW$, $WZ$ and $ZZ$ production. While
    BG exceeds signal at low $m_{cT}$, the largest signal distributions
    are close to the BGs around $m_{cT}\sim 1000$ GeV. Thus, this
    channel might offer a confirming signal to a bulge in the large
    transverse mass distribution from one of the above cases.

In the next two subsections, we attempt to see whether the signals in
the high $p_T$ $h(\to b\bar{b}) + \eslt$ and high $p_T$ $Z(\to
\ell\bar{\ell}) +\eslt$ channels can be further enhanced by requiring
additional soft leptons from the decays of the higgsinos. However,
before turning to this discussion, we mention that we had also attempted
to examine the study the signal from hadronically decaying high $p_T$
$Z$ and $W$ bosons + $\eslt$ events (without any additional soft
leptons) but found that it was hopelessly overwhelmed by the 
background $Z \to \nu\bar{\nu} +$ jet production.

\subsection{$H,\ A\to 3\ell +\eslt$ signal}
\label{ssec:3l}
Here, we attempt to pick out $Z\to\ell\bar{\ell}+\eslt$ events that occur
along with a soft lepton from light higgsino decay. Thus, we 
require: 
\bi
\item Exactly three baseline leptons.
\item At least two leptons satisfy the signal lepton requirement.
\item At least two leptons are OS/SF leptons.
\item At least one OS/SF pair satisfies the $Z$-mass requirement:
  $80\ {\rm GeV}<m(\ell\bar{\ell})<100$ GeV; if more than one pair
  satisfy the $Z$-mass, then the pair closest to $m_Z$ is chosen whilst
  the third is designated $\ell_3$, 
\ei

In addition, we require:
\bi
\item $\eslt|200$ GeV,
\item $\eslt_{,rel}>25$ GeV,
\item $p_T(\ell_3) <30$ GeV,
\item $|\eta (\ell_s)|< 1.5$; $|\eta (\ell\bar{\ell}|<1.3$,
\item $\Delta\phi(\ell\bar{\ell},\eslt )>125^\circ$,
\item $\Delta\phi (\ell_1\ell_2 )<55^\circ$ and
  \item $|\vec{p}_T(Z)+\vec{p}_T(\ell_3\eslt )|>20$ GeV.
    \ei

\begin{figure}[htb!]
\begin{center}
\includegraphics[height=0.4\textheight]{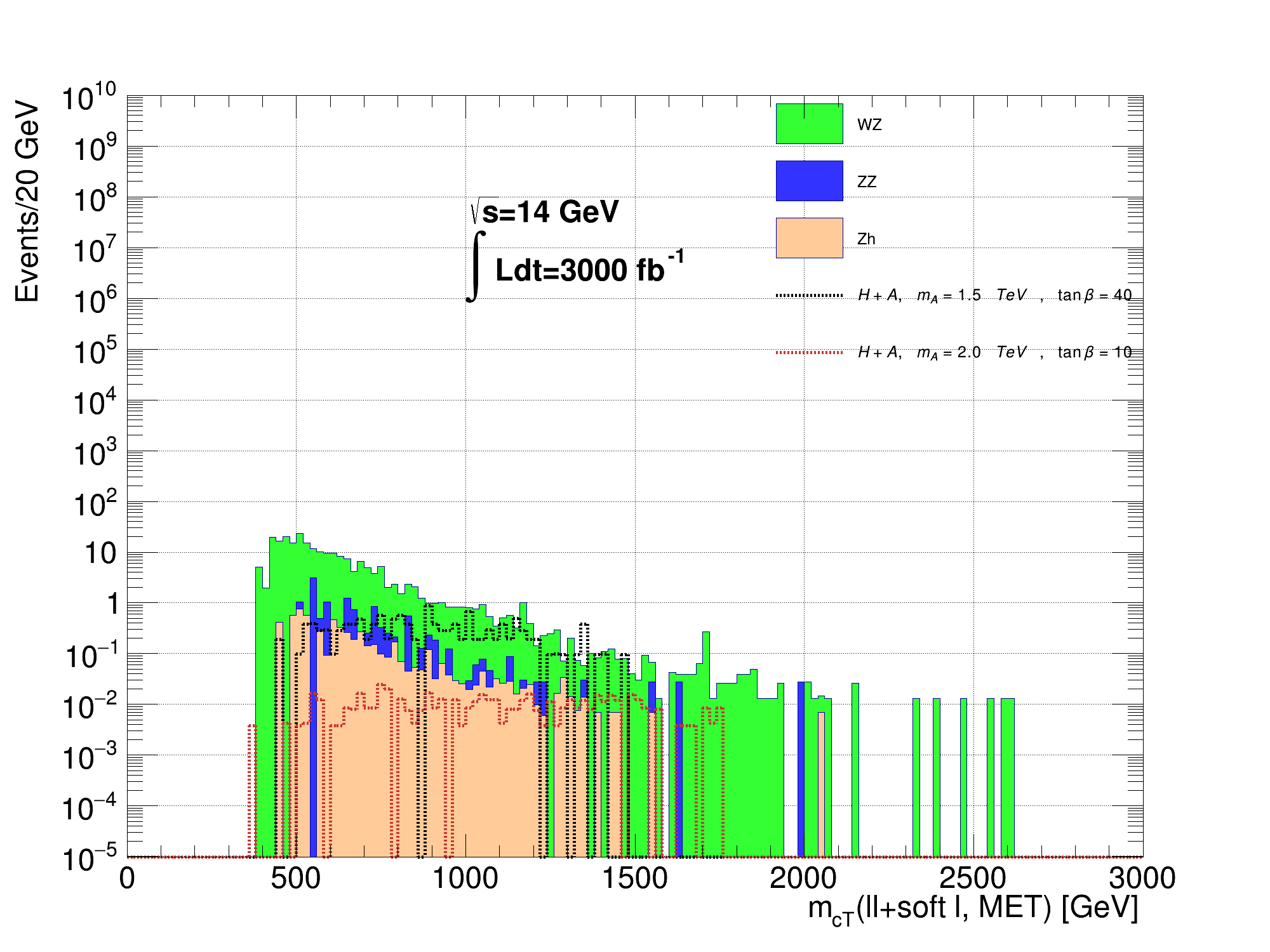}
\caption{Distribution in $m_{cT}(3\ell ,\eslt )$ for
  $pp\to H,A\to 3\ell +\eslt$ events
  at $\sqrt{s}=14$ TeV.
  We show two signal distributions (dashed) along with
  dominant SM backgrounds (not stacked).
\label{fig:3l}}
\end{center}
\end{figure}

    The resultant cluster transverse mass distribution
    $m_{cT}(3\ell,\eslt )$ is shown in Fig. \ref{fig:3l}. We see that,
    as might be expected, the largest SM BG comes from $WZ$ production.
    At very large $m_{cT}(3\ell,\eslt )\agt 1.2$ TeV, our largest signal
    BM point is comparable to SM BG levels.  However, in the best case,
    only a small number of signal events will populate this region.
    Thus, this channel may offer,at best, come corroborative evidence
    for a $H,\ A\to SUSY$ signal.

\subsection{$H,\ A(\to bb)+\ell +\eslt$ signal}
\label{ssec:bbl}

Finally, we attempt to capture an $h\to b\bar{b}+\eslt$ signal where
there is an additional soft lepton from light higgsino decay.  For this,
we require the following.
\bi
\item At least two $b$-jet candidates
\item At least one baseline lepton,
\item Exactly one LRj
\item At least two $b$-jets within the cone of the LRj.
\item Exactly one pair of $b$-jets within the cone of the LRj reconstructs the light Higgs:
  $90\ {\rm GeV}<m(bb)<135$ GeV,
  \item $m(LRj)-m(bb)>-5$ GeV, and
 \item $m(bb)/p_T(LRj)>0.9$.
    \ei
    Then, we also require:
    \bi
  \item $\eslt >225$ GeV,
  \item $\eslt_{,rel}>200$ GeV,
  \item $p_T(\ell )<30$ GeV for any $\ell$ in the event,
  \item $p_T(b_1)>100$ GeV,
  \item $p_T(bb)/p_T(LRj)>0.9$,
  \item $|\eta (bb)|<2$,
  \item $\Delta\phi (bb,\eslt )>145^\circ$,
  \item $max[\Delta\phi(j,LRj)]<150^\circ$ where $j$ cycles over all SR jets in the event, and
  \item $\Delta\phi(b_1,b_2)<65^\circ$.
    \ei

\begin{figure}[htb!]
\begin{center}
\includegraphics[height=0.4\textheight]{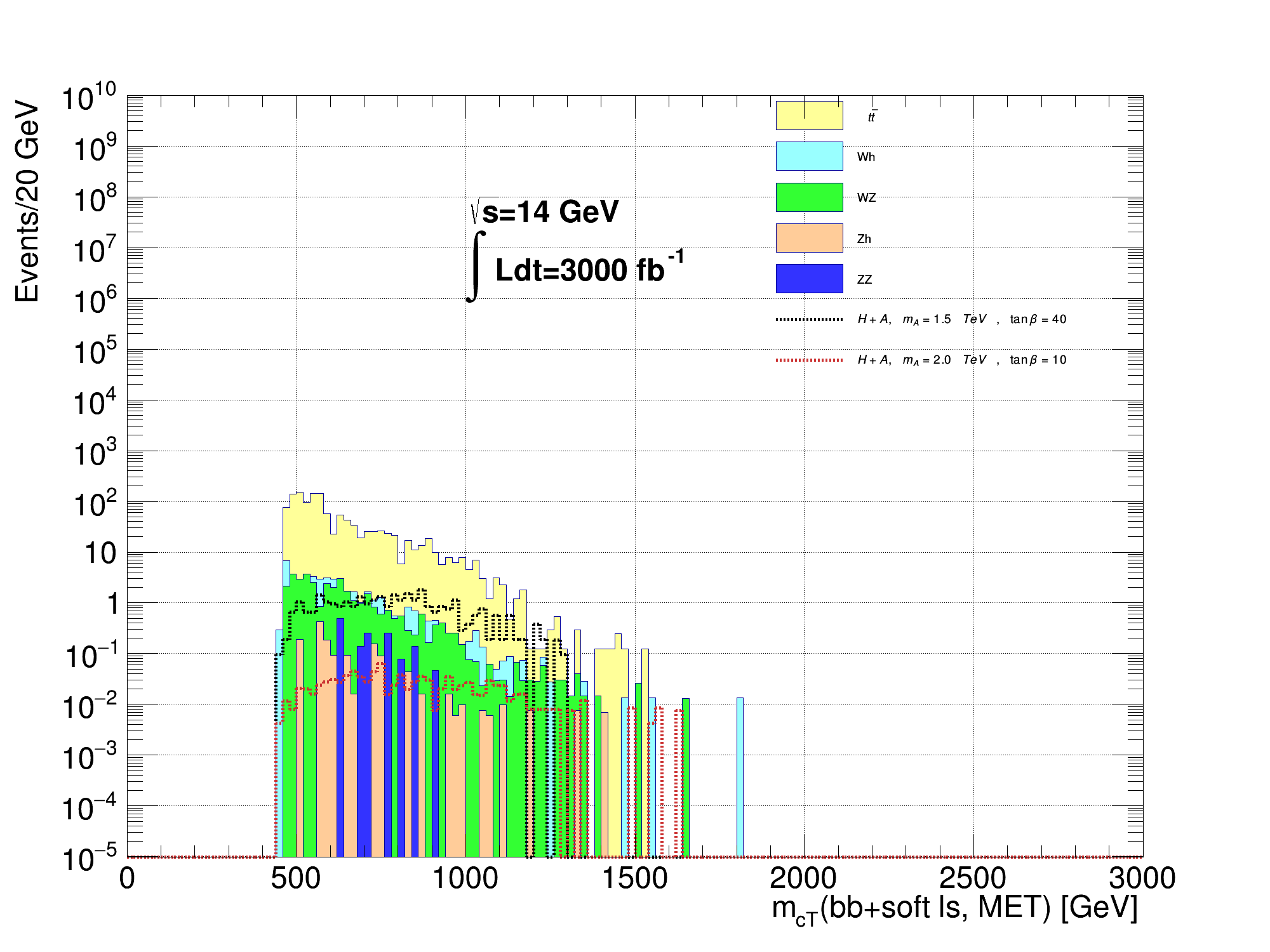}
\caption{Distribution in $m_{cT}(bb\ell ,\eslt )$ for
  $pp\to H,A\to bb\ell +\eslt$ events
  at $\sqrt{s}=14$ TeV.
  We show two signal distributions (dashed) along with
  dominant SM backgrounds (not stacked).
\label{fig:2b1l}}
\end{center}
\end{figure}

    The cluster transverse mass distribution $m_{cT}(bb\ell ,\eslt)$
    distribution is shown in Fig. \ref{fig:2b1l}.  The largest BG comes
    from $t\bar{t}$ production where again the the two $b$-jets
    accidently have $m_{bb} \sim m_h$, with smaller contributions from $Wh$ and
    $WZ$ production. At very large $m_{cT}$ values, signal may emerge
    from BG although the high energy tail is very much rate limited.

\section{Regions of the $m_A$ vs. $\tan\beta$ plane accessible to HL-LHC}
\label{sec:results}

After adopting the above cuts for the various signal channels,
we can now create reach plots in terms of discovery sensitivity or
exclusion limits for $pp\to H,\ A\to SUSY$ in the $m_A$ vs. $\tan\beta$ plane. 
For the discovery plane, we use $5\sigma$ to denote the discovery and
assume the true distribution one observes in experiment corresponds to
signal-plus-background.
Then we test this against the background-only distribution to see if
the background-only hypothesis could be rejected at a $5\sigma$ level.
Specifically, we use the binned transverse mass distributions
from each signal channel as displayed above to obtain the
discovery/exclusion limits.
For the exclusion plane, the upper limits for exclusion of a signal are
set at the 95\% CL and assume the true distribution one observes in
experiment corresponds to background-only.
They are then computed using a modified frequentist $CL_s$
method\cite{Read_2002} with the profile likelihood ratio as the test statistic. 
In both the discovery and exclusion planes, the asymptotic approximation
for getting the median significance is used\cite{Cowan_2011}.
The systematic uncertainty is assumed to be $1\sigma$ of the corresponding
statistical uncertainty, which is a very conservative rule-of-thumb estimate. 

\begin{figure}[htb!]
\begin{center}
  \includegraphics[height=0.2\textheight]{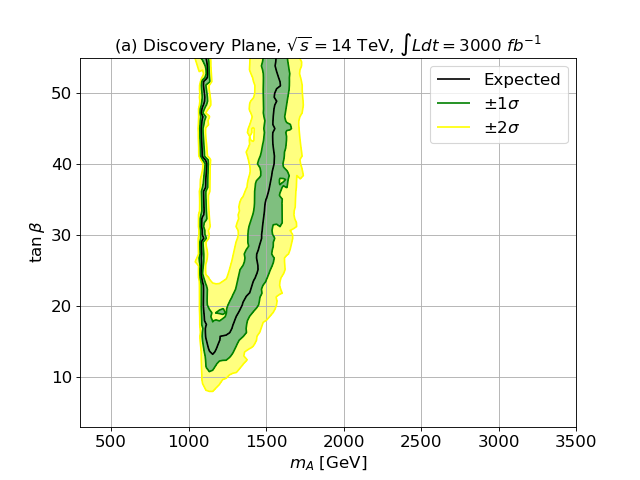}
  \includegraphics[height=0.2\textheight]{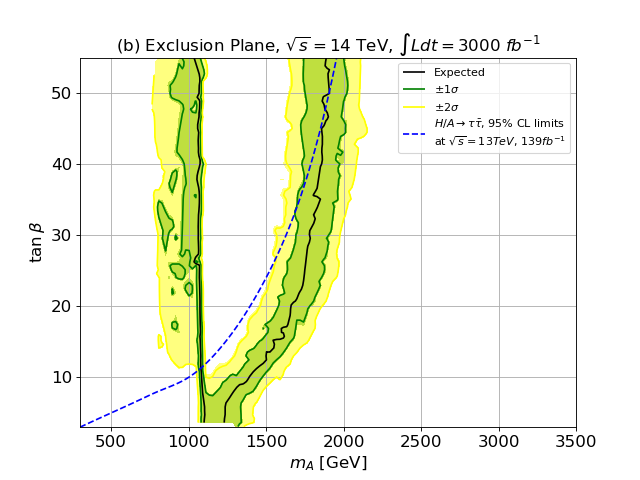}\\
  \includegraphics[height=0.2\textheight]{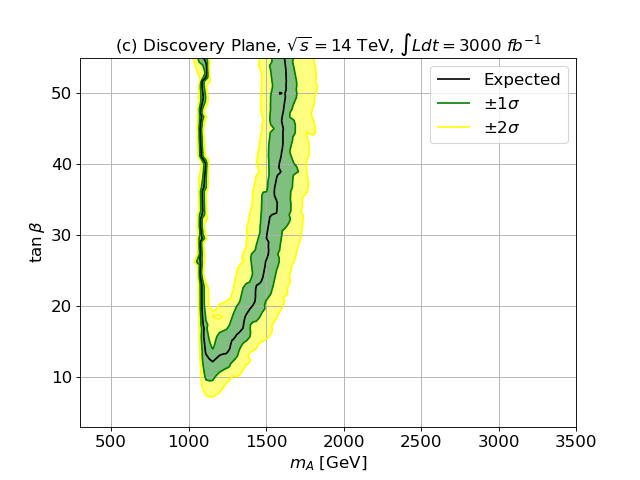}
  \includegraphics[height=0.2\textheight]{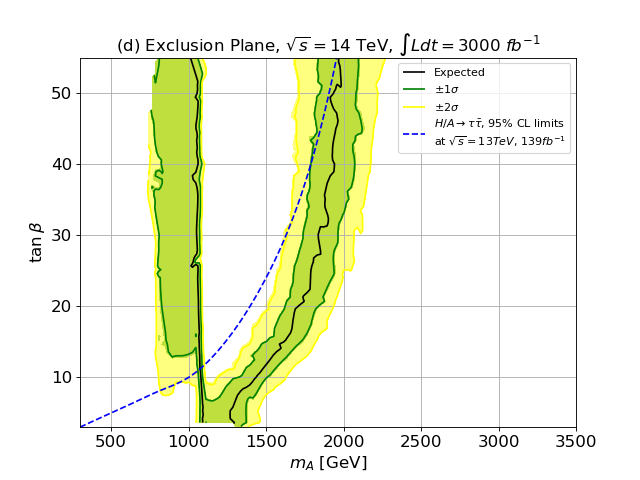}\\
  \includegraphics[height=0.2\textheight]{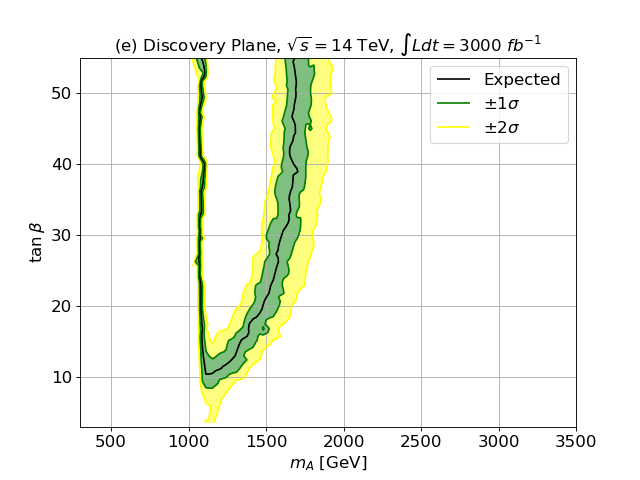}
  \includegraphics[height=0.2\textheight]{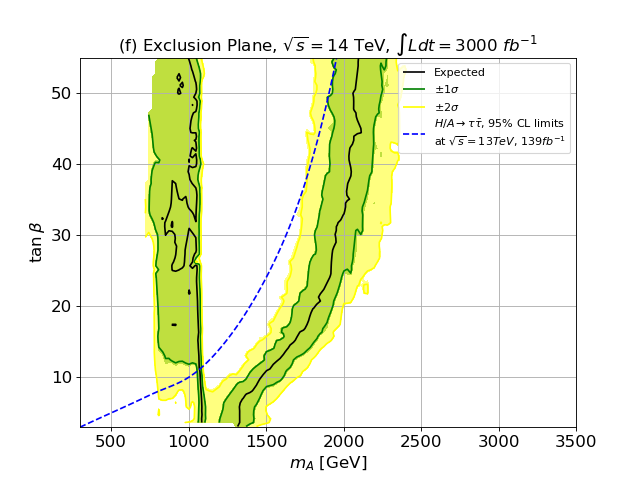}
  \caption{In frame {\it a}), we plot the HL-LHC $5\sigma$ discovery
    contours for $H,A\to bb+\eslt$ events with $\sqrt{s}=14$ TeV and
    3000 fb$^{-1}$.  In frame {\it b}), we plot the corresponding 95\%
    CL exclusion limit.  In frame {\it c}), we plot the $5\sigma$
    discovery reach via the combined $bb+\eslt$ and
    $\ell\bar{\ell}+\eslt$ channels.  In {\it d}), we plot the
    corresponding 95\% CL exclusion.  In {\it e}), we show the $5\sigma$
    contour combining all six discovery channels, while in {\it f}) we
    plot the 95\% CL exlusion limits from the all six channels
    combined. The region above the dashed contour in the frames in the
    right-hand column is excluded at the 95\%CL by ATLAS, albeit in the
    $m_h^{125}$ scenario where the $H$ and $A$ essentially decay only
    via SM modes.
    \label{fig:disc_excl}}
\end{center}
\end{figure}

In Fig. \ref{fig:disc_excl}, we plot our main results for the
discovery/exclusion regions for HL-LHC with $\sqrt{s}=14$ TeV and 3000
fb$^{-1}$ of integrated luminosity in the $m_A$ vs. $\tan\beta$ plane
using our $m_h^{125}({\rm nat})$ benchmark scenario (which is quite
typical for natural SUSY models\cite{Baer:2020kwz}).  In frame {\it a}),
we plot the $5\sigma$ discovery reach using the strongest channel which
is the $H,\ A\to h+\eslt$ with $h\to b\bar{b}$.  The solid black line
denotes the computed reach while the green and yellow bands display the
$\pm 1\sigma$ and $\pm 2\sigma$ uncertainty.  From the plot, we see that
a discovery region does indeed exist, starting around $m_A\sim 1$ TeV
where $H,\ A\to gaugino+higgsino$ begins to turn on.  For this channel,
the discovery region extends out to $m_A\sim 1.5$ TeV for high
$\tan\beta$ where the increasing $H,\ A$ production cross section
compensates for the decreasing $H,\ A\to SUSY$ branching fractions.  The
discovery region pinches off below $\tan\beta\sim 10$ where the $pp\to
H,\ A$ production rates become too small, mainly because the bottom
quark Yukawa coupling becomes small. In frame {\it b}), we plot the 95\%
CL exclusion limit for HL-LHC in the $b\bar{b}+\eslt$ channel. While
this plot has the same low $m_A$ kinematic cutoff, the exclusion limit
now extends out to $m_A\sim 1.85$ TeV for large $\tan\beta\sim 40-50$.
We also see that the exclusion contour extends well below $\tan\beta\sim
10$. The region above the blue dashed contour in the frames in the
right-hand column is excluded at the 95\%CL by the ATLAS search with an
integrated luminosity of 139 fb$^{-1}$ \cite{ATLAS:2020zms} for the
signal from $H,A \to \tau\bar{\tau}$ decays, albeit in the $m_h^{125}$
scenario where the $H$ and $A$ essentially decay only via SM modes. If
SUSY decay decays of $H$ and $A$ are important, these will reduce the
branching fractions for the decays to tau pairs, and the allowed region
to the right of the blue controur will be somewhat larger.

In Fig. \ref{fig:disc_excl}{\it c}), we plot the $5\sigma$ discovery
contour, but this time we combine the two strongest channels:
$b\bar{b}+\eslt$ and $\ell^+\ell^-+\eslt$.
The added channel from $H,\ A\to Z+\eslt$ has a similar lower cutoff
(as expected) around $m_A\sim 1$ TeV but now extends to $m_A\sim 1.65$ TeV
for high $\tan\beta$-- an increase in discovery reach by $\sim 150$ GeV
from the frame {\it a}) result. The corresponding 95\% CL exclusion
from the two combined channels is shown in frame {\it d}), where the high
$\tan\beta$ limit now extends out to $m_A\sim 2$ TeV, again a gain in
$m_A$ reach of $\sim 150$ GeV.

In frame {\it e}), we show the HL-LHC $5\sigma$ discovery reach contour
which is gained by combining all {\it six} signal channels from
Sec. \ref{sec:channels}. In this case, the discovery reach starts at the
same kinematic cutoff, but now extends as high as $m_A\sim 1.75$ TeV, a
gain in reach of 250 GeV over the single-channel results from frame {\it
  a}).  In frame {\it f}), we plot the corresponding 95\% CL exclusion
contour from the combined six signal channels. For this case, the
contour now extends to $m_A\sim 2.15$ TeV at high $\tan\beta$; this is a
gain in $m_A$ reach of $\sim 300$ GeV over the single channel exclusion
results from frame {\it b}).  Meanwhile, the exclusion contour extends
well below $\tan\beta\sim 10$ in the lower portion of the plot.

To illustrate how the significance increases with the increasing number
of signal channels which are included, we list in Table \ref{tab:signif}
our calculated significances for our $\tan\beta =40$, $m_A=1.5$ TeV BM
point, assuming 3000 fb$^{-1}$ of integrated luminosity. Our main
purpose here is to illustrate the relative contributions of each SUSY
channel to the significance. For this BM case with just the
$b\bar{b}+\eslt$ channel, we already have a significance of 5.53. As we
include more signal channels, the significance climbs to over $7\sigma$
for this particularly favourable BM point.
\begin{table}[h!]
\centering
\begin{tabular}{cc}
\hline
signal channel & Significance \\
$b\bar{b}$ & 5.53\\
$b\bar{b}+2\ell$ & 6.25 \\
$b\bar{b}+2\ell+ \ell LRj$ & 6.92 \\
$b\bar{b}+2\ell+\ell LRJ+ \ell 2b$ & 7.14 \\
$b\bar{b}+2\ell+\ell LRJ+ \ell 2b+3\ell$ & 7.47 \\
$b\bar{b}+2\ell+\ell LRJ+ \ell 2b+3\ell+1\ell$ & 7.54 \\
\hline
\end{tabular}
\caption{Significance of the signal for six different signal
  channels for our BM point with $\tan\beta =40$ and $m_A=1.5$ TeV
  at HL-LHC with 3000 fb$^{-1}$.
}
\label{tab:signif}
\end{table}

\section{Conclusions}
\label{sec:conclude}

In this paper, we have studied the prospects of HL-LHC to detect the
heavy neutral Higgs bosons of natural SUSY models via their decays into
SUSY particles. Natural SUSY models may be regarded as the most plausible
of SUSY scenarios since they naturally explain why the weak scale lies
in the $m_{W,Z,h}\sim 100$ GeV range whilst the SUSY breaking scale is in the
(multi-) TeV range. It has been argued that
SUSY models with radiatively-driven
naturalness\cite{Baer:2012up,Baer:2012cf} are 
expected as most likely to emerge from string landscape statistics;
the landscape then actually 
predicts $m_h\sim 125$ GeV with sparticles beyond present LHC search
limits\cite{Baer:2017uvn}.

For this class of models, with light higgsinos and TeV-scale gauginos,
heavy neutral Higgs bosons decay dominantly to $gaugino+higgsino$ once
these modes are kinematically allowed, provided $\tan\beta$ is not very
large.  The SUSY decay modes diminish the usually-assumed SM decay modes
(such as $H,\ A\to \tau\bar{\tau}$) while opening new discovery
possibilities. Here, we identify the most promising discovery channels
as $W+ \eslt$, $Z+ \eslt$ and $h+\eslt$, where the $W,Z$ or $h$ come from the
decay of the heavy gaugino daughter. We proposed sets of cuts designed
to optimize extraction of signal from background. While $W+\eslt$ is
beset with huge SM backgrounds, mainly from direct $W$ production, the
$h(\to b\bar{b})+\eslt$ and $Z(\to \ell\bar{\ell})+\eslt$ and ($Z$ or
$h$)$\to \tau\bar{\tau} + \eslt$ channels are much more promising. We
also examined the possibility that the signal would be enhanced by
requiring additional (soft) leptons from the decay of the higgsinos.

Specifically, in each of these channels, we analysed various binned
transverse mass distributions for signal plus background to test against
the background only hypothesis to obtain $5\sigma$ discovery contours
and 95\% CL exclusion contours  in the $m_A$ vs. $\tan\beta$ plane at
the HL LHC.
Our main result is shown in Fig.~\ref{fig:disc_excl} where we show the
reach plots for 1. the strongest channel, $h+\eslt$ where
$h\to b\bar{b}$, 2. this channel combined with the next strongest $Z(\to
\ell\bar{\ell})$ channel, and 3. for all six $H,\ A\to SUSY$ discovery
channels. The $H,\ A\to SUSY$ signal can occur at viable levels for
$m_A\sim 1-2.1$ TeV for large $\tan\beta\agt 30$ with lower exclusion
contours for lower $\tan\beta$ where $H,\ A$ production rates become
much smaller.  We can compare our reach for $H,\ A\to SUSY$ discovery
channels to the recently computed discovery reach via the
$H,\ A\to\tau\bar{\tau}$ mode as presented in Ref. \cite{Baer:2022qqr}.
In that work, it was found that even with SUSY decay modes allowed in
the same $m_h^{125}({\rm nat})$ scenario, the discovery/exclusion
contours extend well past the $H,\ A\to SUSY$ modes which were
investigated here.  This is perhaps due to the advantages of
$m(\tau\bar{\tau})$ reconstructing a resonance around
$m(\tau\bar{\tau})\sim m_A$ and in general lower SM background levels
for the $\tau\bar{\tau}$ channel. It would be very instructive to
examine the impact of the SUSY decays of the heavy Higgs bosons at
future hadron colliders such as FCChh where the larger center-of-mass
energy allows would allow for the production of Higgs bosons with masses
of several TeV and kinematically unsuppressed decays to SUSY
particles. 

{\it Acknowledgements:} 

This material is based upon work supported by the U.S. Department of Energy, 
Office of Science, Office of High Energy Physics under Award Number DE-SC-0009956 and DE-SC-0017647.


\bibliography{HAsusy}
\bibliographystyle{elsarticle-num}

\end{document}